\documentstyle[sprocl,epsf]{article}      
\newcommand{\postscript}[2] {\setlength{\epsfxsize}{#2\hsize}
\centerline{\epsfbox{#1}}}
\begin{document}
\title
{Trapped Bose-Einstein Condensed Gas with Two and Three-Atom
Interactions \footnote{
Contribution to the 
{\it Internationl Workshop on Collective Excitations in Fermi
and Bose Systems}, Serra Negra, Brazil, September 14-17, 1998. 
To be published by World Scientific, Singapore.}}
\author{A. Gammal$^{(a)}$, T. Frederico$^{(b)}$ and L. Tomio$^{(a)}$}
\address{
$^{(a)}$
Instituto de F\'\i sica Te\'orica, Universidade Estadual Paulista,\\  
01405-900 S\~{a}o Paulo, Brazil \\
$^{(b)}$
Departamento de F\'{\i}sica, Instituto Tecnol\'ogico da Aeron\'autica,\\ 
Centro T\'ecnico Aeroespacial, 12228-900 S\~ao Jos\'e dos Campos,
SP, Brazil
}
\maketitle
 
\begin{abstract}
The stability of a Bose-Einstein condensed state of trapped ultra-cold atoms
is investigated under the assumption of an attractive two-body and 
a repulsive three-body interaction. 
The Ginzburg-Pitaevskii-Gross (GPG) nonlinear Schr\"odinger equation is
extended  to include an effective potential dependent on the square of the
density and solved numerically for the $s-$wave.
The lowest frequency of the collective mode is determined through the 
Fourier transform of the time dependent solution and
its dependences on the number of atoms and 
the strength of the three-body force are studied.
We show that the addition of three-body dynamics can allow the number 
of condensed atoms to increase considerably, 
even when the strength of the three-body force is very small
compared with the strength of the two-body force.
\end{abstract}
\vskip 0.5cm

Recently, the theoretical research on Bose-Einstein condensation (BEC),
a phenomenon predicted by Einstein more than 70 years ago, received
considerable support from the experimental evidences of BEC in
magnetically trapped  weakly interacting 
atoms~(\cite{and96,mew96,brad97}).
The nature of the effective atom-atom interaction determines the
stability of the condensed state:  the two-body pseudopotential 
is repulsive for a positive $s-$wave atom-atom scattering
length and 
it is attractive for a negative scattering length~(\cite{huang}). 
The ultra-cold trapped atoms with repulsive two-body interaction undergoes 
a phase-transition to a stable Bose condensed state, in a number of cases
found experimentally, as for $^{87}$Rb~(\cite{and96}), for
$^{23}$Na~(\cite{mew96}) and $^7$Li~(\cite{brad97}). \ 
However, a condensed state of atoms with negative $s-$wave
atom-atom
scattering length would be unstable unless the number of atoms is small
enough such that the stabilizing force provided by the harmonic
confinement in the trap overcomes the attractive interaction, as found on 
theoretical grounds~(\cite{rup95,baym96}). 
It was indeed observed in the $^7$Li gas~(\cite{brad97}), for which the
$s-$wave scattering length is $a=(-14.5 \pm 0.4)$ \AA, that the  number
of 
allowed atoms in the Bose condensed state was limited to a maximum value
between 650 and 1300, which is consistent with the mean-field  
prediction~(\cite{rup95}). 
An earlier experiment~(\cite{brad95}) suggested that the number of atoms
in 
the condensate state was significantly larger than the theoretical
predictions with two-body pseudopotential. This is consistent with an
addition of a potential derived from three-body interaction, which
can extend considerably the region of stability for the condensate
even for a very small strength of the three-body force.

Expanding our universe of possible effective interactions, it was
reported in Ref.~(\cite{esry}) that a sufficiently dilute 
and cold Bose gas exhibits similar three-body dynamics for both signs of 
the $s-$wave atom-atom scattering length. They concluded that the
long-range 
three-body interaction between neutral atoms is effectively repulsive for
either sign of the scattering length. Supposing that a repulsive
three-body effective interaction is present in the atomic system, one can
imagine that  a stable condensate will be formed in the trap for a number
of bosons large enough, such that the three-body interaction  overcomes the
attraction of the two-body interaction~(\cite{josse}).

In the present work, we investigate the competition between the attractive
two-body interaction, originated from a negative two-atom $s-$wave
scattering length, and a repulsive three-body effective interaction. 
(The latter can be originated by the existence of a weakly bound 
three-boson state, as it will be explained.)
We show that, in a dilute gas, a small repulsive three-body force
added to an attractive two-body interaction
is able to stabilize the condensate beyond the critical number of atoms
in the trap, found just with attractive two-body force~(\cite{rup95}).
The Ginzburg-Pitaevskii-Gross  nonlinear Schr\"odinger
equation~(\cite{gin})
is extended  to include an effective potential dependent on the square of
the density and solved numerically for $s-$wave.
The stability is studied using a weak perturbation probe and the time
dependent equation is solved with the ground state as the starting point.
The lowest frequency of the collective mode is determined through the
Fourier transform of the time dependent solution and
its dependences on the number of atoms and the strength of the
three-body force are studied.
In this case, a signature for a repulsive three body force 
can appear with the possibility of two stable solutions,
for a fixed number of atoms, even for a very small repulsive
three body interaction.
Also, in this case, one stable solution is found for a large number of 
atoms, beyond the maximum expected with only two-body interaction.

In the following, we present the formalism, where the original 
Ginzburg-Pitaevskii-Gross (GPG) non-linear equation~(\cite{gin}), 
which includes a term proportional to the density (two-body interaction), 
is extended through the addition of a term proportional to the 
squared-density (three-body interaction).
Next, after reducing such equation to dimensionless units, we study 
 numerically the $s-$wave solution by varying the corresponding 
dimensionless parameters, which are related to the two-body scattering 
length, the strength of the three-body interaction
and the number of atoms in the condensed state. 
As particularly observed in Ref.~(\cite{hs}), 
to incorporate all two-body scattering processes in 
such many particle system, the two-body potential should be replaced by
the many-body $T-$matrix. \ \ Usually, at very low energies, this is
approximated by the two-body scattering matrix, which is directly
proportional to the scattering length~(\cite{baym96}). So, to obtain
the desired equation, we first consider the effective Lagrangian, 
which describes the condensed wave-function in the Hartree approximation,
implying in the Ginzburg-Pitaevskii-Gross (GPG) energy
functional~(\cite{gin}):
\begin{eqnarray}
{\cal {L}}&=&\int d^3r \left[ \frac{i\hbar}2\Psi^{\dagger}(\vec r)
\frac{\partial \Psi(\vec r)}{\partial t}-\frac{i\hbar}2\frac{\partial 
\Psi^{\dagger}(\vec r)}{\partial t}\Psi(\vec r)+\frac{
\hbar ^2}{2m}\Psi^{\dagger}(\vec r)\nabla ^2\Psi(\vec r)
\right. \nonumber \\ &&\left.
- \frac m2 \omega^2 r^2 |\Psi(\vec r)|^2\right] +{\cal{L}}
_{\rm{I}}\ .  \label{lag}
\end{eqnarray}
In our description, the atomic trap is given by a rotationally symmetric
harmonic potential, with angular frequency $\omega$, and 
${\cal{L}}_{\rm{I}}$ gives the effective atom interactions up to three 
particles.

The effective interaction Lagrangian for ultra-low temperature bosonic 
atoms, including two and three-body scattering at zero energy, is 
written as: 
\begin{eqnarray}
&&{{\cal L}}_{\rm I}= -\frac{1}{2}\int 
d^3r_1d^3r_2d^3r^{\prime}_1d^3r^{\prime}_2 
\Psi^\dagger 
(\vec{r^{\prime}}_1)\Psi^\dagger (\vec{r^{\prime}}_2) \Psi (\vec{r}_1) 
\Psi(\vec{r}_2)
\\
&&\times \left\langle \vec{r^{\prime}}_{12} \left| T^{(2)}(0) \right| 
\vec{r}_{12} \right\rangle 
\delta^3(\vec{r}_1^\prime+\vec{r}_2^\prime-\vec{r}_1-\vec{r}_2)
\nonumber \\
&&-\frac{1}{3{\rm{!}}}\int
d^3r_1d^3r_2d^3r_3d^3r^{\prime}_1d^3r^{\prime}_2d^3r^{\prime}_3 
\Psi ^{\dagger} (\vec{r}_1^{\prime}) \Psi ^{\dagger} (\vec{r}_2^{\prime}) 
\Psi^\dagger (\vec{r}_3^{\prime}) \Psi (\vec{r}_1)\Psi 
(\vec{r}_2)\Psi (\vec{r}_3) 
\nonumber \\
&&\times
\left\langle\vec{r}_{12}^{\prime}\vec{R}_3^{\prime}\left|T^{(3)}(0)-
\sum_{j<k}T^{(2)}_{jk} (0- K_i)\right|\vec{r}_{12}\vec{R}_3\right\rangle 
\delta^3 ( \vec{R}_{123}^{\prime}-\vec{R}_{123}
) \ ,  \label{li}
\nonumber
\end{eqnarray}
where $\vec r_{12}$ and $\vec R_3$ are the relative coordinates, given by
$\vec r_{12}=\vec r_1-\vec r_2$ and 
$\vec R_3= \vec r_3-(\vec r_1+\vec r_2)/2$; and 
$\vec R_{123}\equiv (\vec r_1+\vec r_2+\vec r_3)$. \ \
$T^{(3)}(0)$ and $T^{(2)}_{jk}(0)$ are the corresponding 
three-body $T-$matrix and two-body $T-$matrix for the pair ${jk}$, which
are evaluated at zero-energy.
The two-body $T-$matrix for each pair $(jk)$ is subtracted from 
$T^{(3)}(0)$ to avoid double counting and 
$K_i$ is the kinetic energy operator for particle $i$. \ 

We can approximate the above effective interaction Lagrangian at 
low densities by averaging the $T-$matrices over the relative coordinates,
considering that the thermal wave-length is much greater than the 
characteristic interaction distances. 
\begin{eqnarray}
&&{\cal{L}}_{\rm I}=-\frac{1}{2}\int d^3r^{\prime}_{12}d^3r_{12}
\left\langle \vec{r^{\prime}}_{12} \left| T^{(2)}(0) \right| \vec{r}_{12}
\right\rangle \int d^3r 
\left|\Psi (\vec{r})\right|^4  
\nonumber \\
&-&\frac{1}{3 {\rm{!}}}\int d^3r^{\prime}_{12} d^3R^{\prime}_3d^3r_{12}
d^3R_3 \left\langle \vec{r^{\prime}}_{12}\vec{R^{\prime}}_3 \left|
T^{(3)}(0)-\sum_{j<k}T^{(2)}_{jk}(0- K_i) \right| \vec{r}_{12}\vec{R}_3
\right\rangle  
\nonumber \\
&\times & \int d^3r 
\left|\Psi (\vec{r})\right|^6 \ .  \label{li1}
\end{eqnarray}
The integrations of the $T$-matrices over the relative coordinates gives the
zero momentum matrix elements: 
\begin{eqnarray}
\int d^3r^{\prime}_{12}d^3r_{12} \left\langle \vec{r^{\prime}}_{12}\left|
T^{(2)}(0)\right|\vec{r}_{12}\right\rangle &=& 
(2 \pi)^3 \left\langle \vec p_{12}=0 \left| T^{(2)}(0)\right|\vec{p}_{12}=0
\right\rangle \nonumber \\
&=& \frac{4\pi\hbar^2 a}{m} \ ,  \label{t2}
\end{eqnarray}
where $a$ is the two-body scattering length. For the connected three-body
$T-$matrix, we have 
\begin{eqnarray}
&&\int d^3r^{\prime}_{12} d^3R^{\prime}_3d^3r_{12} d^3R_3 \left\langle
\vec{
r^{\prime}}_{12}\vec{R^{\prime}}_3 \left|
T^{(3)}(0)-\sum_{j<k}T^{(2)}_{jk}(0- K_i) \right| \vec{r}_{12}\vec{R}_3
\right\rangle  \nonumber \\
&&= (2 \pi)^6 \left\langle \vec{p}_{12}=0;\vec{P}_3=0 \left|
T^{(3)}(0)-\sum_{j<k}T^{(2)}_{jk}(0- K_i) \right| \vec{p}_{12}=0;\vec{P}_3=0
\right\rangle \nonumber \\
&&= 2 \lambda_3 \ ,  \label{t3}
\end{eqnarray}
where $\lambda_3$ is the  strength of the three-body effective interaction.

The nonlinear Schr\"odinger equation, which describes the condensed
wave-function in the mean-field approximation, is
obtained from the effective Lagrangian given in Eq.~(\ref{lag}).
By considering the interaction in Eq.~(\ref{li1}) and the
definitions in Eqs.~(\ref{t2}) and (\ref{t3}), it can be written
as~(\cite{fw})
\begin{equation}
\left[ -\frac{\hbar ^2}{2m}{\bf \nabla }^2+\frac m2\omega ^2
r^2 -\frac{4\pi \hbar ^2\ |a|}m|\Psi (
\vec r)|^2+\lambda_3|\Psi (\vec r )|^4\right] \Psi (\vec r)\ =\ \mu \Psi
(\vec r)\ ,  \label{sch}
\end{equation}
where $\mu $ is the chemical potential, fixed by the number $N$
of atoms in the condensed state: 
\begin{equation}
\int d^3r|\Psi (\vec r)|^2\ =\ N\ .
\label{norm1}
\end{equation}
The physical scales present in Eq.(\ref{sch}) can be easily recognized 
by working with dimensionless equations. By rescaling Eq.(\ref{sch}) for 
the $s-$wave solution, we obtain
\begin{equation}
\left[ -\frac{d^2}{dx^2}+\frac14 x^2 -\frac{|\Phi (x)|^2}{x^2}
+g_3\frac{|\Phi (x)|^4}{x^4}\right] 
\Phi (x)\ =\ \beta \Phi (x)\ ,  \label{schd}
\end{equation}
where $x\equiv \sqrt{{2 m \omega}/{\hbar}} \ r$ and
$ \Phi(x)\equiv\sqrt{8 \pi |a|}\; r\Psi (\vec r)$.
The dimensionless parameters, related to the chemical potential and
the three-body strength are, respectively, given by
\begin{equation}
\beta\equiv \frac{\mu}{\hbar \omega}\;\;\;{\rm and}\;\;\; 
g_3\equiv\lambda_3 \hbar \omega \left[\frac{m}{4 \pi\hbar^2a}\right]^2; 
\label{beta-g}
\end{equation}
and the normalization for $\Phi (x)$, obtained from Eq.~(\ref{norm1}), 
defines a number $n$ related to the number of atoms $N$:
\begin{equation}
\int^{\infty}_0 dx|\Phi (x)|^2\ = n , \;\;\; {\rm{where}} 
\;\;\;
n\equiv 2 N |a|\sqrt{\frac{2m\omega}{\hbar}} \ .
\label{norm2}
\end{equation}
The boundary conditions in Eq.(\ref{schd}) are given by~(\cite{rup95})
\begin{equation}
\Phi(0)=0 \;\;\; {\rm and} \;\;\; \Phi(\infty)\sim
C\exp{\left[-x^2/4+(\beta-1/2)\ln(x)\right]}\label{boundary} .
\end{equation}

The numerical solutions of Eq.~(\ref{schd}) are obtained for several
values of $\beta$, using three values of $g_3$ to characterize the 
solutions. 
We have used the Runge-Kutta (RK) and ``shooting" method
to obtain the corresponding solutions in each case.
The stability assignment was made by studying the corresponding time 
dependent Scr\"{o}dinger equation, using the Crank-Nicolson (CN) method
[see Ref.~(\cite{Clark})]. 
The numerical procedure to determine such stability was done in the
following way: when applying the CN method, we started by using the static
solution obtained from the RK method and observed if the modulus of
the wave function remained constant. If this was occuring for a
long period of time (of about 500 units of dimesionless time $\tau =
\omega t$) the solution was considered stable, otherwise unstable. 
In another procedure we added a weak perturbation to the potencial. We
examined the time evolution of a selected point of the wave-function.
The lowest collective oscilations ($\omega _{col}$) were determined by
using the Fourier transform of such result~(\cite{Clark}).

Figures 1-4 presents our main results.

{\small{ 
\postscript{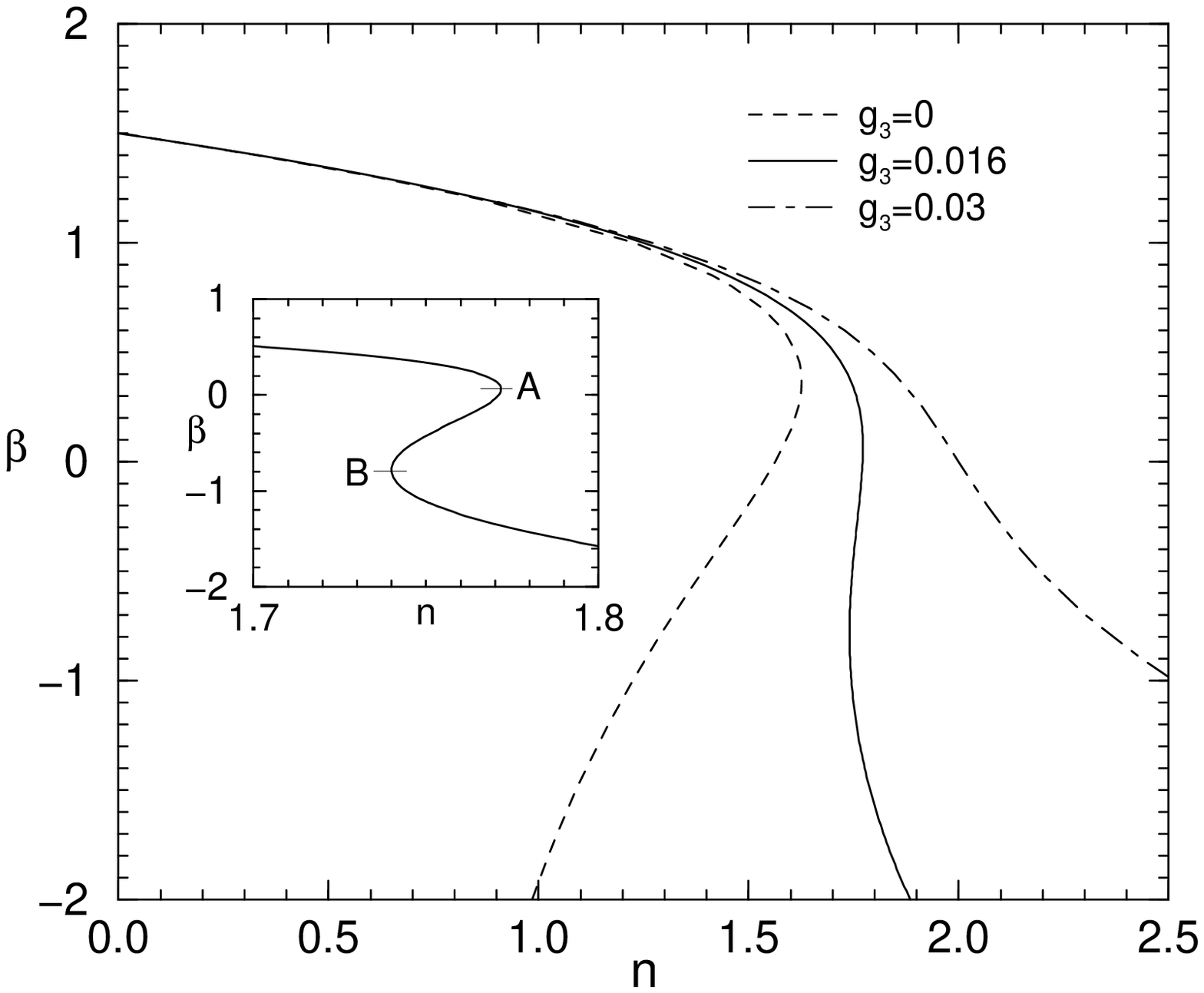}{0.6}    

{\bf Fig. 1 }  
The chemical potential $(\beta )$ in units of $\hbar \omega$ as a function of
$n$ for three values of the non-dimensional strength of the 
repulsive three-body effective interaction $(g_3)$. $g_3=0$ (dashed
line); $g_3=0.016$ (solid line); $g_3=0.03$ (dot-dashed line)
[see Eqs.~(\ref{beta-g}) and (\ref{norm2})]. 
In the inset, for $g_3 = 0.016$, the solutions linking A and B
are unstable.
\vskip 0.2cm

\postscript{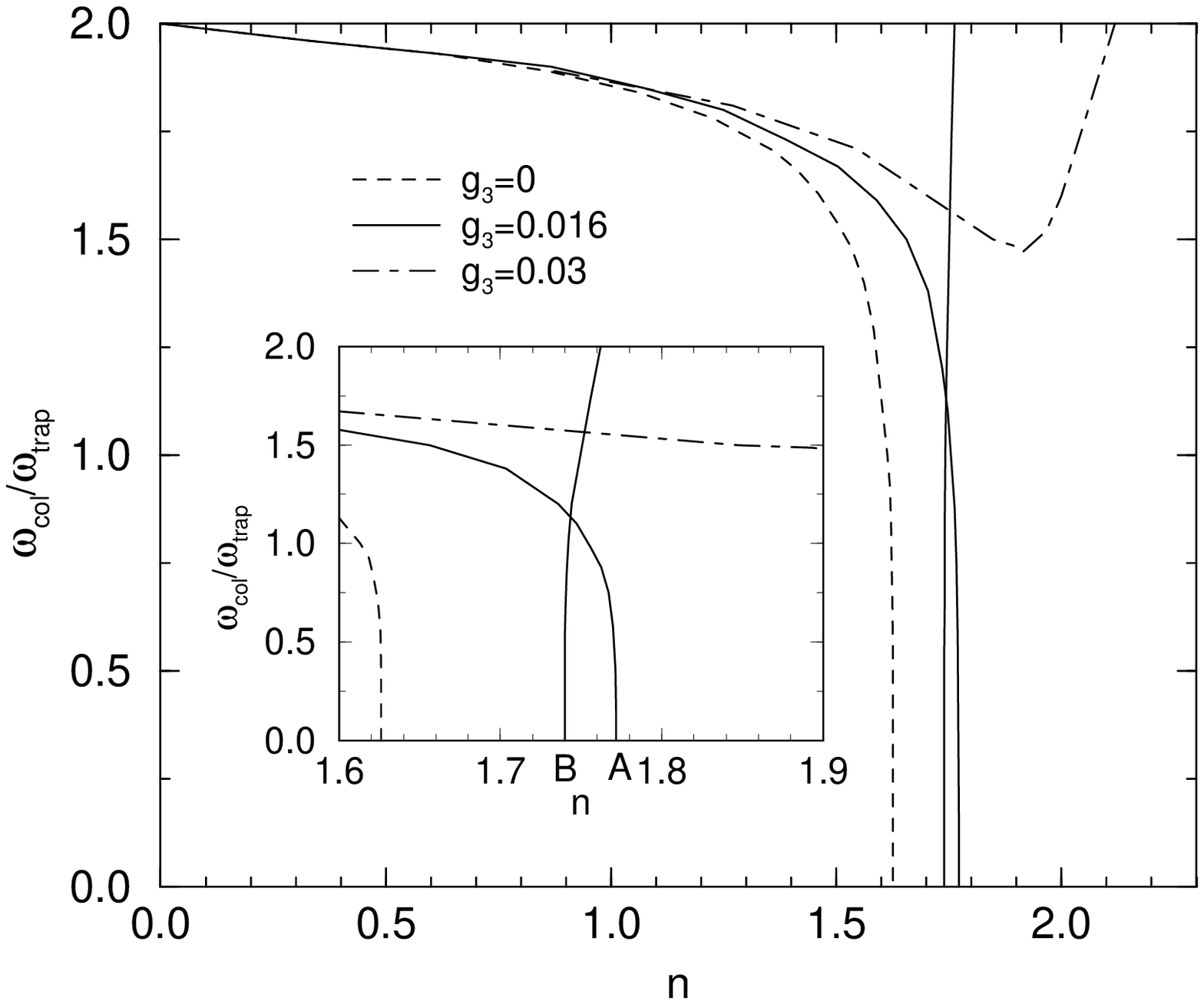}{0.6}    

{\bf Fig. 2 }  
Frequencies of the lowest collective excitations (with $l=0$) as a
function of $n$ for three values of the dimensionless strength $g_3$ 
The points A and B correspond to the points shown in the inset of Fig.1,
where the frequencies go to zero.

  
\postscript{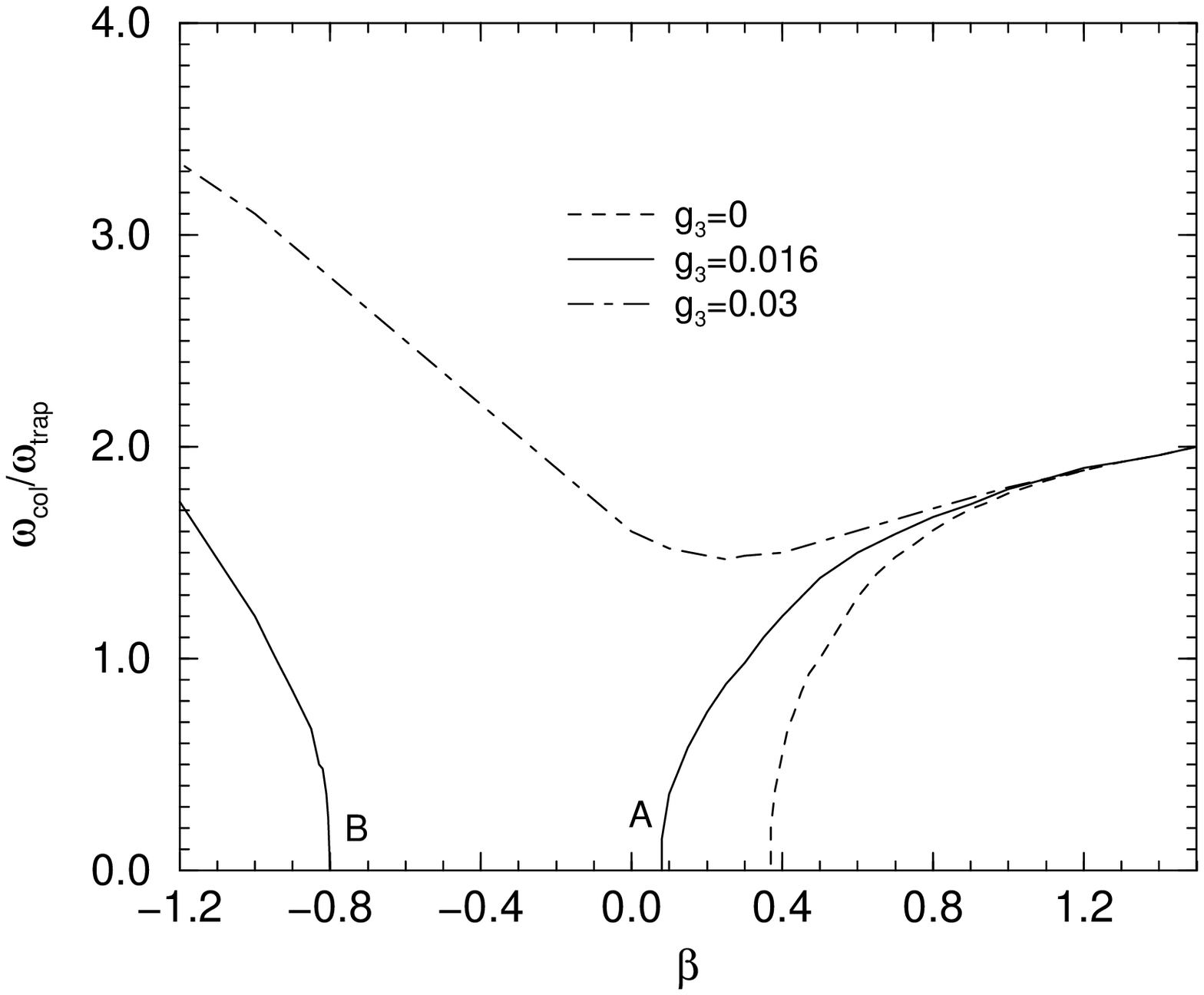}{0.6}    

{\bf Fig. 3 }  
Frequencies of the lowest collective excitations ($l=0$), as a function of
$\beta$ for three values of the nondimensional strength $g_3$. 
The points A and B correspond to the points shown in the inset of Fig.1,
where the frequencies go to zero.

\vskip 0.5cm 
  
\postscript{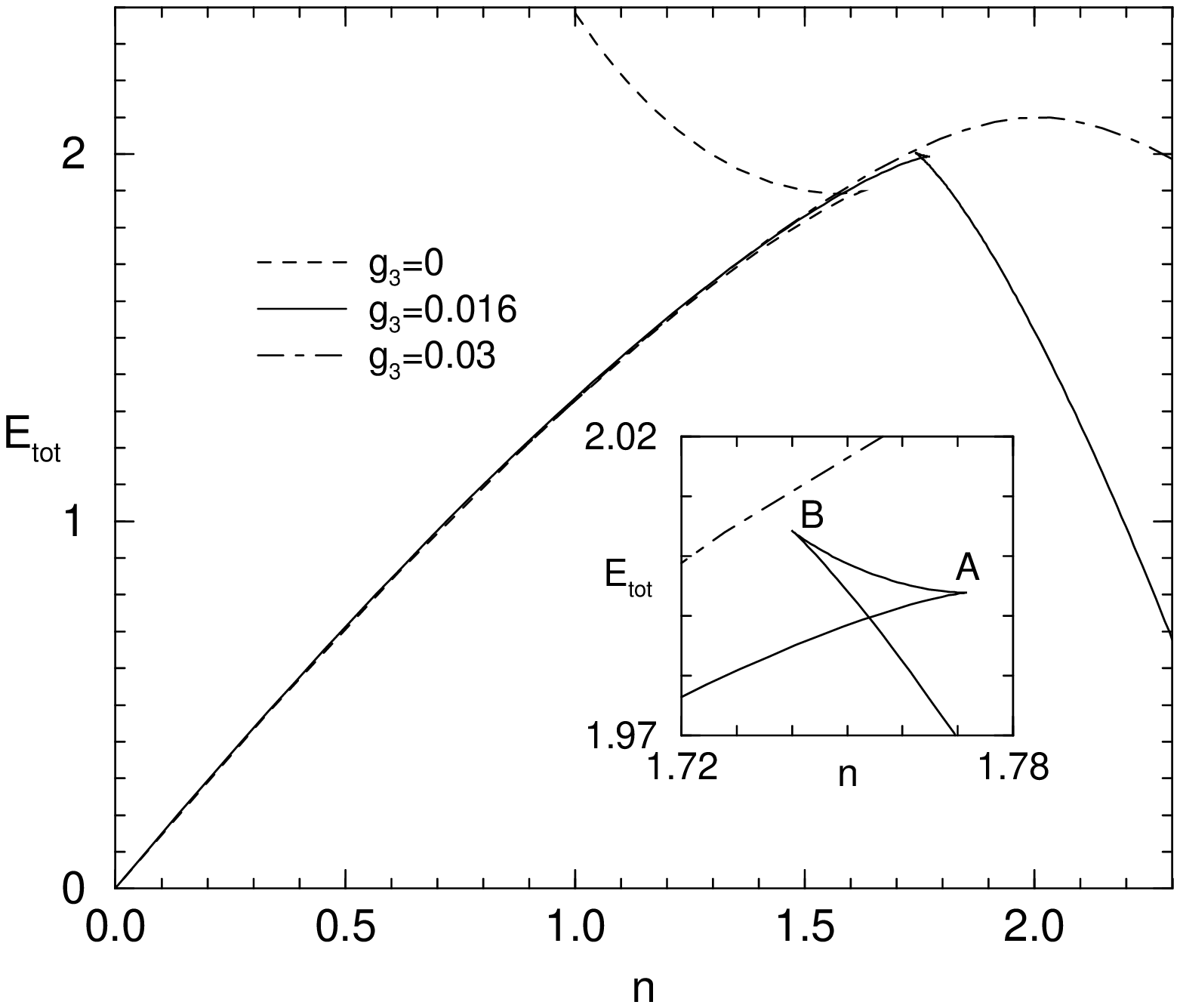}{0.6}    

{\bf Fig. 4 }  
Total energy of the Bose condensate, 
given in units of $\hbar\omega/\left(2|a|\sqrt{(2m\omega/\hbar}\right)$,
is shown as a function of $n$, for three values of the nondimensional 
strength $g_3$. The line linking A and B in the inset correspond to 
the unstable region shown in Fig.~1.
 }}
\newpage
In Figure 1 we present $\beta $ as function of $n$ for $g_{3}$ equal to
0, 0.016 and 0.03. \ Our calculation for $g_{3}=0$ agrees with the
result presented in Ref.~(\cite{rup95}), with the maximum number of
atoms limited to $n_{max}\approx 1.62$~\footnote{
Our $n$ is equal to $|C_{nl}^{3D}|$ of Ref.~(\cite{rup95}).}.
As we can see, for $n \le n_{max}$
two solutions are possible, the lower being unstable. 
\ For $g_{3}$ slightly higher than 0 a new pattern appears, as
one can see for $g_3=0.016$ presented in the figure. 
We can divide the figure, for $g_3=0.016$ (for example), in
three regions; the first, stable, goes from $\beta =1.5$ until point A
(presented in the inset); an unstable region appears from the
point A till B; another stable region goes from B towards $n\to\infty$. 

Figure 2 shows the collective frequencies $\omega _{col}$ as function of 
$n$ for the first mode ($l=0$). \ The solutions corresponding to $g_{3}=0$
agree
well with the ones given in Ref.~(\cite{SR});
and they loose stability as $\omega _{col}\rightarrow 0.$  
By this criteria, we determined the regions of stability for
$g_{3}=0.016$.  For $g_{3}=0.03$ all the solutions are stable. 
The collective frequencies are also shown as a function of $\beta $ 
in Figure 3.
The total energy is shown in Figure 4.
The case $g_{3}=0$ reproduces the results already found by
Houbiers and Stoof~(\cite{hs}), including the unstable (higher)
solutions. For  $g_{3}=0.016$ the part of the plot linking points A and 
B is unstable, corresponding to the unstable region of Figure 1.
Otherwise it is stable. Finally, for $g_{3}=0.03$, the function of the
energy in terms of $n$ is always single valued and stable.


The relevance of the three-body interaction was never emphasized in BEC 
studies. \ However, in 1985, it was pointed out in Ref.~(\cite{kagan}) 
that 
an easier experimental approach to probe density fluctuations is to
consider an observable directly sensitive to the probability
of finding three atoms near each other, which will correspond to the
loss rate of atoms due to three-body recombination.
Such a three-body recombination rate in BEC, was considered recently in 
Refs.~(\cite{fedi}), (\cite{burt}) and (\cite{kagan0}) 
It was shown in Ref.~(\cite{fedi}) that the three-body recombination
coefficient of ultracold atoms to a weakly bound $s$ level goes to 
infinity in the Efimov limit~(\cite{efimov}).  
The Efimov limit is a particularly interesting three-body effect,
which happens when the two-body scattering length is very large 
(positive or negative).
In this case, with the two boson energy close to zero,
the three-boson system presents an increasing  number of loosely 
bound three body states, which have large spatial extension and do not
depend on the details of the interaction~(\cite{3atom}). 
In such limit, when the three-body binding is close to zero, the
strength $\lambda_3$ can be large enough to give a sensible 
value for $g_3$. So, our main motivation here was to provide 
an extension to the GPG equation~(\cite{gin}), which considers a
three-body interaction and, in this way, provides the 
framework for a numerical investigation of the relevance
of three-body interaction in Bose-Einstein condensation.
In this letter, we presented results for 
the chemical potential, $\mu$,  the
number of atoms in the condensed state ($N$), and frequency of the
lowest collective mode
for a  range of values of $\lambda_3$,  expressed respectively
in terms of the dimensionless parameters  $\beta$, $n$, 
$\omega_{col}/\omega_{trap}$ and $g_3$.
The determination of $\lambda_3$ has to be considered in particular 
three-boson systems.

To summarize, our calculation presents, at the mean-field level,
the consequences of a repulsive three-body effective
potential added to an attractive two-body interaction
for the Bose condensed atomic state. We propose
a general framework for discussing  such phenomenon. 
As we have shown with the present approach, a slightly repulsive 
three body interaction provides stable solutions for higher number of
atoms where their existence is forbidden when considering only two body
potential. \ This can be a signature of the presence of an effective
repulsive three body interaction in Bose condensates.
\vskip 1cm

{\bf Acknowledgments}

This work was presented at the {\it Internationl Workshop
on Collective Excitations in Fermi and Bose Systems}, Serra Negra,
Brazil, September 14-17, 1998 (to be published by World Scientific, Singapore,
1999). Work partially supported by Funda\c c\~ao de Amparo \`a
Pesquisa do Estado de S\~ao Paulo and Conselho Nacional de Desenvolvimento
Cient\'\i fico e Tecnol\'ogico.


\begin{thebibliography}{99} 
\bibitem{and96} M.R. Andrews, M.-O. Mewes, N.J. van Druten, D.S.
Durfee, D.M. Kurn, W.Ketterle, Science {\bf 273}, 84 (1996);
M.H. Anderson, J.R. Ensher, M.R. Matthews, C.E.
Wieman, E.A. Cornell, Science {\bf 269}, 198 (1995).
\bibitem{mew96} M.-O. Mewes,M.R. Andrews, N.J. van Druten, D.M. Kurn,
D.S. Durfee, and W.Ketterle, Phys. Rev. Lett. {\bf 77}, 416 (1996).
\bibitem{brad97} C.C. Bradley, C.A. Sackett and R.G. Hulet, 
Phys. Rev. Lett. {\bf 78}, 985 (1997);
C.C. Bradley, C.A. Sackett, J.J. Tollet and R.G. Hulet,  
Phys. Rev. Lett. {\bf 79}, 1170 (1997).
\bibitem{huang}  K. Huang, {\it Statistical Mechanics}, (John Wiley and
Sons, New York, 1987).
\bibitem{rup95} M. Edwards and K. Burnett, Phys. Rev. {\bf A51}, 1382
(1995); P.A. Ruprecht, M.J. Holland, K. Burnett, and M. Edwards, Phys. Rev.
{\bf A51}, 4704 (1995). 
\bibitem{baym96}  G. Baym and C.J. Pethick, Phys. Rev. Lett., {\bf 76}, 6
(1996).
\bibitem{brad95} C.C. Bradley, C.A. Sackett, J.J. Tollet and R.G. Hulet, 
Phys. Rev. Lett. {\bf 75}, 1687 (1995).
\bibitem{esry}  B.D. Esry, C.H. Greene, Y. Zhou, and C.D. Lin, J. Phys. B  
{\bf 29}, L51 (1996). 
\bibitem{josse} C. Josserand and S. Rica, Phys. Rev. Lett. {\bf 78}, 1215
(1997).
\bibitem{gin}  V.L. Ginzburg and L.P. Pitaevskii, Zh. Eksp. Teor. Fiz. 34,
1240(1958) [Sov. Phys. JETP 7, 858 (1958)]; L.P. Pitaevskii, Sov. Phys. JETP
13,451 (1961); E.P. Gross, J. Math. Phys. 4, 195 (1963).  
\bibitem{hs} M. Houbiers and H.T.C. Stoof, Phys. Rev. A {\bf 54}, 5055
(1996).
\bibitem{fw} A.L. Fetter and J.D. Walecka, {\it Quantum Theory of Many -
Particle Systems} (McGraw-Hill, New York, 1971).
\bibitem{Clark} P.A. Ruprecht, M. Edwards, K. Burnett, and C.W.Clark,
Phys. Rev. A {\bf 54}, 4178 (1996). 
\bibitem{SR} K.G. Singh and D.S. Rokhsar, Phys. Rev. Lett. {\bf 77},
1667 (1996).
\bibitem{kagan} Yu. Kagan, B.V. Svistunov, and G.V. Shlyapnikov, 
JETP Lett. {\bf 42}, 209 (1985).
\bibitem{fedi}  P.O. Fedichev, M.W. Reynolds, and G.V. Shlyapnikov, Phys. 
Rev. Lett. {\bf 77}, 2921 (1996). 
\bibitem{burt} E.A. Burt, R.W. Ghrist, C.J. Myatt, M.J. Holland, 
E.A. Cornell, and C.E. Wieman, Phys. Rev. Lett. {\bf 79}, 337 (1997).
\bibitem{kagan0} Yu. Kagan, A.E. Muryshev, and G.V. Shlyapnikov, 
Phys. Rev. Lett. {\bf 81}, 933 (1998).
\bibitem{efimov}  V. Efimov, Phys. Lett. {\bf B 33}, 563 (1970); 
Comm. Nucl. Part. Phys. {\bf 19}, 271 (1990). 
\bibitem{3atom} A.E.A. Amorim, T. Frederico, and L. Tomio, Phys. Rev. C
{\bf 56}, 4 (1997).
\end{thebibliography}
\end{document}